\documentclass[11pt]{article}   	
\usepackage{geometry}                		
\usepackage{graphicx}				
\usepackage{amssymb,amsmath}
\usepackage{gensymb} 
\usepackage{hyperref} 
\hypersetup{	
colorlinks=true, 
linkcolor= blue,	
citecolor=blue,	
} 
\usepackage{subfigure, color}
\usepackage[dvipsnames]{xcolor} 
\usepackage{tikz}

\usepackage{cite}

\usepackage{gensymb} 
\usepackage{subfigure, color, xcolor,bm}

\title{Enhanced flight performance 
in non-uniformly flexible wings}
\author{Lionel VINCENT$^1$, Min ZHENG$^1$, John H. COSTELLO$^{2,3}$, and Eva KANSO$^1$\footnote{Corresponding author: Kanso@usc.edu} \\[1ex]
\small{${}^1$ Aerospace \& Mechanical Engineering, University of Southern California, Los Angeles, CA 90089} \\
\small{${}^2$ Biology Department, Providence College, Providence, RI 02918 USA} \\
\small{${}^3$ Whitman Center, Marine Biological Laboratory, Woods Hole, 02543 MA}}
\date{}							

\begin{document}

\maketitle

\begin{abstract}
\small{The flexibility of biological propulsors such as wings and fins is believed to contribute to the higher performance of flying and swimming animals compared with their engineered peers. Flexibility seems to follow a universal design rule that induces bending patterns at about one-third from the distal tip of the propulsor's span. However,  the aerodynamic mechanisms that shaped this convergent design and the potential improvement in performance are not well understood.  Here we analyze the effect of heterogenous flexibility on the flight performance of tumbling wings. Using experiments, numerical simulations, and scaling analysis, we demonstrate that spanwise tip flexibility that follows this empirical rule leads to improved flight performance. Our findings attribute this improvement to flutter-induced drag reduction. This mechanism is independent of the wing's auto-rotation and represents a more general trait of wings with non-uniform tip flexibility. We conclude by analyzing the effects of both spanwise and chordwise non-uniformity on the flight performance of tumbling seedpods.}
\end{abstract}


\section*{Introduction}

Flight and swimming in the animal kingdom involves the periodic motion of appendages providing thrust and lift. Unlike their engineered counterparts, most animal propulsory appendages are flexible \cite{LucasCostello2014}. In a recent study, flexibility was found to follow a potentially universal rule: flexibility occurs at about two-third of the propulsor's span \cite{LucasCostello2014}. Although the reason of this well-defined ratio is not yet known, it has been hypothesized that bending has beneficial effects on the aerodynamics of the appendage.

\begin{figure}[!t]
   \centering
   \includegraphics[width=0.4\textwidth]{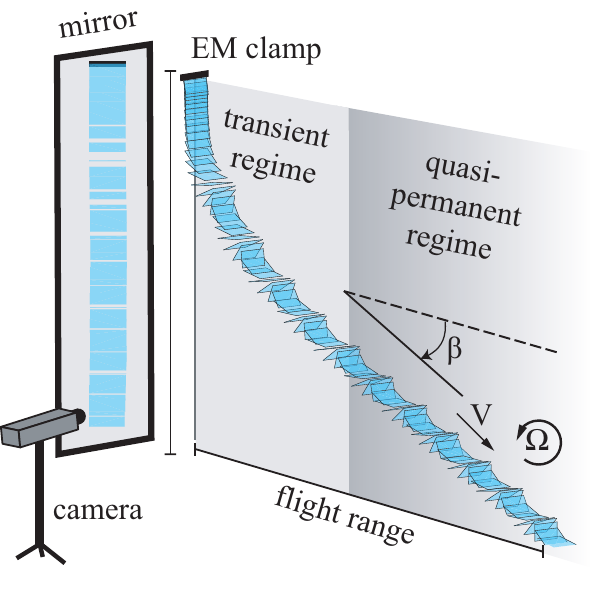} 
   \caption{\footnotesize 
   Tumbling wings: spanwise and chordwise non-uniform wings are released edge-on with zero initial conditions using an electromagnetic clamp and reach a quasi-steady tumbling behavior.}
    \label{fw_fig:xp_setup}
  \end{figure}

Owing to the difficulty of performing in-vivo studies, mechanical models are typically used to assess the effect of flexibility on propulsion. Propulsion efficiency in undulatory swimming is traditionally studied by means of a heaving and pitching flexible panels  \cite{Alben2008,Alben2012,Quinn2015,Heathcote2007,Michelin2009,RaspaRamananarivo2014} or wing-shaped airfoils \cite{Ramananarivo2010}, both with uniform flexibility; for these systems, it has been found that performance improvement requires a careful selection of the excitation parameters and stiffness. Similarly, heaving and pitching panels with heterogeneous flexibility show conditional improvement compared to rigid panels \cite{Lucas2015}. Rather than passive propulsors, several studies rely on prescribed kinematics \cite{Young2009,Le2010,Mittal2006}; again, it is found that improvement in efficiency  depends on the details of the appendage's time-dependent shape. A common ground in these previous studies is the focus on \emph{chordwise} deformation. There have been rare attempts, with conflicting results, to assess the influence of \emph{spanwise} flexibility on performance \cite{LiuBose1997,Kang2011,Heathcote2008}.

A simpler yet related biological system, serving as a model system for the aerodynamics of passive flight, and with potential to shed light on the more complex problem of animal flight problem, is seed dispersal. Winged seedpods that tumble or gyrate are known to generate lift forces that prolong their airborne phase~\cite{Lentink2009}.  Such seedpods can be found in a wide variety of shapes, sizes, and flexibility properties. Historical studies of wings tumbling end-over-end about their spanwise axis focused on rigid wings \cite{Dupleich1941,Belmonte1998,Mahadevan1999}, providing predictions for the falling velocity, tumbling rate, and mapping the boundaries of the tumbling regime. More recent contributions gave a detailed account of their time-dependent dynamics \cite{PesaventoWang2004,AndersenWang2005}. In the only study known to the authors involving flexible tumbling wings, uniform flexibility was found to be unconditionally detrimental to tumbling flight \cite{TamBushKudrolli2010}; the detrimental effect was attributed to an inertial buckling instability, akin to the Euler buckling instability of elastic beams, in which aerodynamic effects are subdued to inertial and elastic forces \cite{TamBushKudrolli2010}; the wing buckled and tumbled in its buckled state much like a rigid curved wing. In contrast, a later study showed that flexibility of freely fluttering plates could lead to an increase in flight performance \cite{Tam2015}.

\begin{table*}[!t]
\begin{center}
\includegraphics[width=0.9\textwidth]{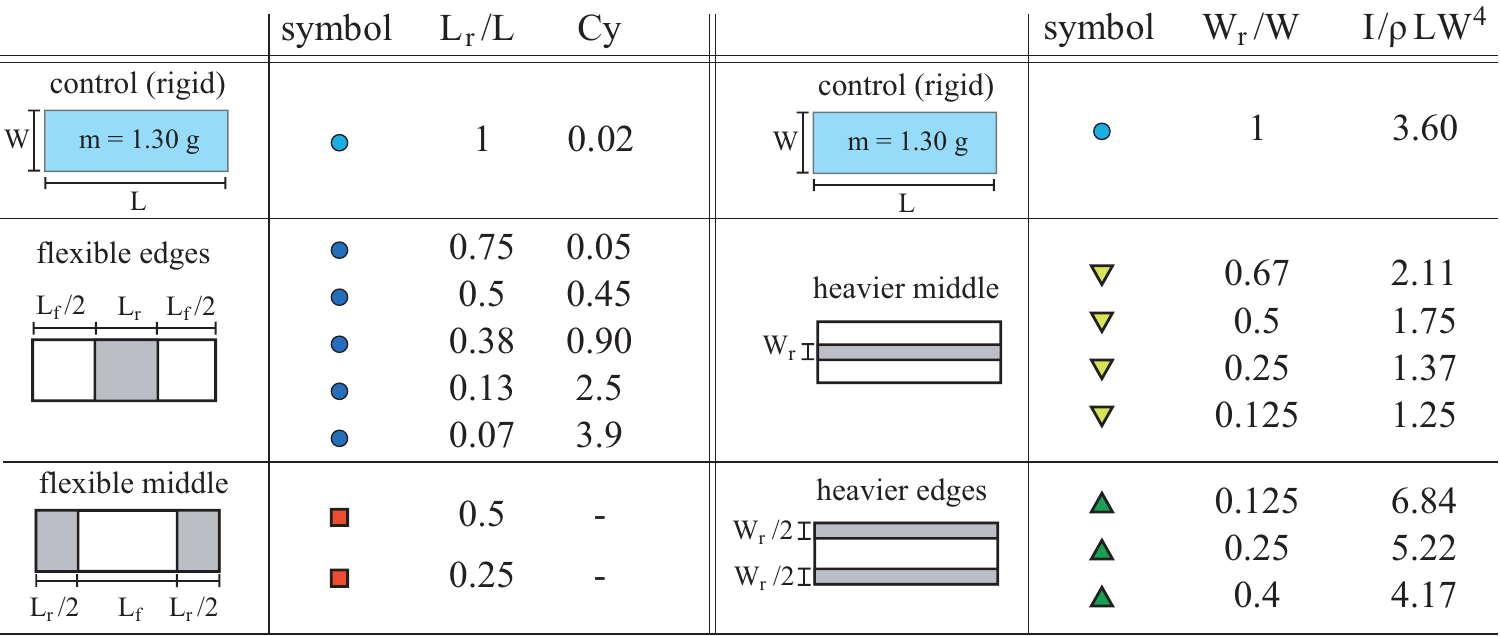}
\end{center}
\caption{Control wing and two families of wings with the same mass $m=1.30$ g, planform $L = 12$ cm and $W = 4.2$ cm, but with  non-uniform mass distribution in the spanwise direction (left column) and in the chordwise direction (right column). Wings with  spanwise non-uniformity have the same dimensionless moment of inertia $I/\rho L W^4 = 3.60$ about the tumbling axis, while wings with chordwise non-uniformity have the same Cauchy number $\textrm{Cy} = 0.02$, defined as the ratio of  aerodynamic to elastic forces. As the wing stiffness increases, $\textrm{Cy}$ decreases and the wing approaches a rigid wing.}
\label{fw_table:prop_wings}
\end{table*}

In this work, we used passively tumbling flight as a model system for exploring the effect of non-uniform wing designs on the flight performance in a regime where aerodynamic effects are dominant. Particularly, we explored  the behavior of wings with non-uniform mass distribution in the spanwise direction first, then in the chordwise direction. The first design exhibited bending patterns that were excitable during tumbling and that led to improved flight under conditions consistent with the empirical observations in~\cite{LucasCostello2014}. The second showed no bending dynamics but the mass non-uniformity influenced the moment of inertia of the wings, which in turn affected the flight performance in non-trivial ways.


\section*{Experimental Design}

\paragraph{Wing design.}
We designed two families of rectangular wings of non-uniform mass distribution but constant total mass $m = 1.3 \pm 0.05$ g and geometry $L \times W = 12 \times4.2$ mm.   
All wings were constructed from three identical, flexible, sheets of paper of thickness $h_f = 0.1$ mm, density $\rho_s = 780$ Kg/m$^3$, and Young's modulus $E=3.6$ Gpa, as measured from dynamic bending tests (see Supplementary Document). A uniform (control) wing was constructed by uniformly overlaying all three sheets of paper together to obtain a wing of overall thickness $h = 0.3$ mm.

Non-uniform wings were constructed as follows:
starting with a single flexible sheet, discrete regions either at the outer or middle portions of this sheet were sandwiched between additional layers of paper composed from the remaining two sheets. 
These regions are highlighted in grey in Table~\ref{fw_table:prop_wings}.
In the left column, we describe wings with non-uniform mass distribution in the spanwise direction.  The portions of the wings made of multiple sheets of paper have length $L_r$ and thickness $h_r$. Mass conservation implies that $h L = h_f L_f + h_r L_r$, where $L = L_f + L_r$ and $L_f$ is the length of the flexible portion of the wing.
The right column of Table~\ref{fw_table:prop_wings} shows the geometric properties of the wings with chordwise mass distribution. The portions composed of multiple sheets of paper are also indicated in grey and have width $W_r$.

We computed the Cauchy number $\textrm{Cy}=\rho  L_f^3 V_g^2 / 16 B$, defined as the non-dimensional ratio of the pressure force $(\rho U^2/2)(L_f/2)^3$ produced by a flow of speed $V_g$ on the wing and the bending rigidity $B=E W h_{f}^3 / 12$ of the wing; see Table~\ref{fw_table:prop_wings}. Here, we assumed a characteristic velocity equal to the terminal velocity of a freely falling wing $V_g = \sqrt{2\rho_s g h/\rho}$, where $\rho = 1.20$~Kg/m$^3$ is the density of air ($\rho_s \gg \rho$). The Cauchy number characterizes the deformation of an elastic solid under the effect of flow. 
Larger $\textrm{Cy}$ implies increased flexibility. Wings with spanwise mass non-uniformity exhibited larger flexibility, while the control wing and the chordwise non-uniform wings were mostly rigid. We also computed the moment of inertia of these wings about the spanwise axis of rotation.  The moment of inertia remained constant for the wings with spanwise heterogeneity but varied by nearly one order of magnitude for the chordwise heterogeneous wings.


\paragraph{Flight experiments.} We released each wing, edge-on with zero initial velocity, from an electromagnetic clamp (Figure~\ref{fw_fig:xp_setup}). All wings settled into an auto-rotating (tumbling) motion at quasi-constant rotation rate, in a regime well beyond the flutter-to-tumbling transition~\cite{Belmonte1998}. Trajectories were captured using a high-speed camera at 400 frames per second. A mirror was positioned so that the camera simultaneously recorded  the side and back views of the wing. Using an in-house image processing algorithm, we extracted the quasi-permanent rotational velocity $\Omega$, the translational velocity $V$, and the angle of descent $\beta$ between the wing's trajectory and the horizontal direction (Figure~\ref{fw_fig:xp_setup}). The values of  $\Omega$, $V$, and $\beta$ were independent of variations in initial conditions, consistent with previous findings~\cite{AndersenWang2005,Mahadevan1999}; {the values obtained were within the range: $\Omega = 25$-$50$~rad/s, $V = 1.5$-$1.9$ m/s, and $\beta = 38\degree$-$70\degree$}.

\section*{Results and Discussion}


\begin{figure*}[!t]
   \centering
   \includegraphics[width=\textwidth]{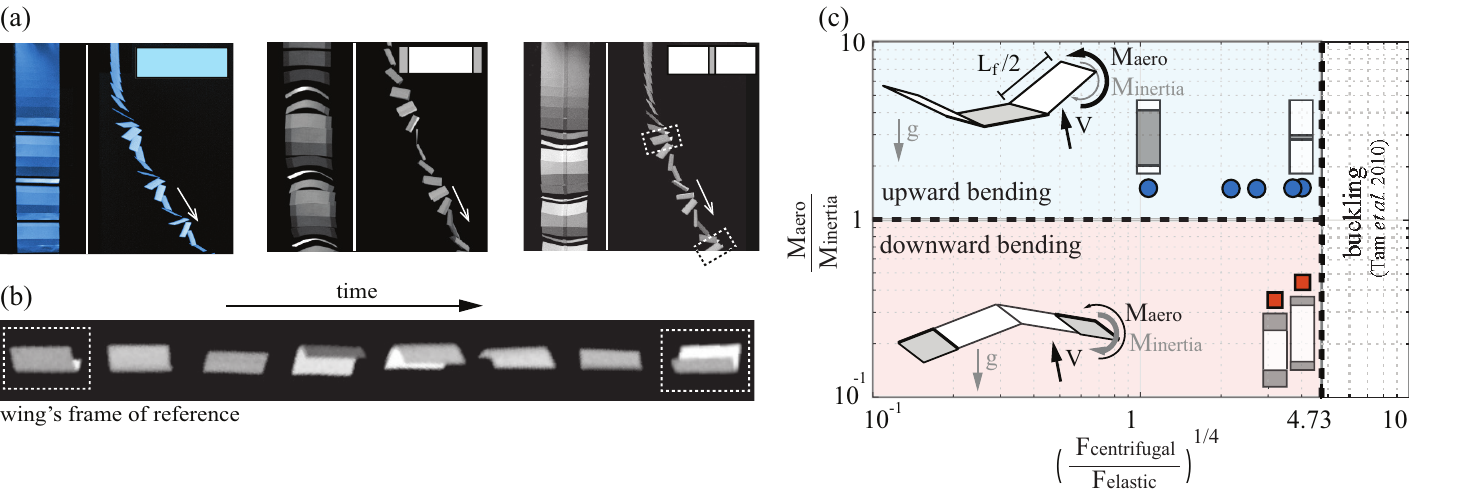} 
   \caption{\footnotesize (a) Control wing tumbles without bending, non-uniform  wings with flexible middle  bend downwards while wings with flexible outer edges bend upwards. (b) The non-uniform wings flap back and forth in their own frame of reference. (c) The downward or upward  bending depends on on the ratio of aerodynamic-to-inertial loads.  Transition to buckling at a locked bent configuration occurs when centrifugal forces are dominant \cite{TamBushKudrolli2010},  in a regime beyond the parameter values considered here.  
   }
   \label{fw_fig:parameters_and_patterns}
\end{figure*}

\paragraph{Spanwise flexible wings.} 
The control wing tumbled without bending while the non-uniform wings bent during flight, as evident from the back view of the wings in Figure~\ref{fw_fig:parameters_and_patterns}(a).
That is to say, the control wing behaved as a rigid wing under the aerodynamic, gravitational and inertial loads applied here, while the non-uniform wings did not.
Wings with flexible base and rigid edges flew `tip first', displaying a `$\cap$' shape, while wings with rigid middle sections and flexible edges flew `base first', displaying a `$\cup$' shape when viewed in the frame of reference of the lab. Both exhibited flapping-like behavior when viewed in the wing's own frame of reference (Figure~\ref{fw_fig:parameters_and_patterns}b).  This behavior differs from the tumbling motion of wings with uniform flexibility, which were reported to buckle under centrifugal forces and rotate around their spanwise axis at a locked, curved configuration,~\cite{TamBushKudrolli2010}.   In fact, for all the wings we considered, the ratio of centrifugal to elastic forces was below the critical value delineating the buckling instability~\cite{TamBushKudrolli2010}.


\paragraph{Scaling analysis.} The upward or downward bending of the spanwise non-uniform wings is due to an imbalance between the aerodynamic and inertial moments acting on the deformable portion of the wing. To verify this statement,
we estimated the forces and moments that cause bending using a scaling argument.  Maximum bending occurred when the wing was in its horizontal configuration.
At this instant, aerodynamic forces act directly upwards and the gravitational and inertial forces directly downwards, leading to upward bending ($\cup$ shape) when the former are dominant and downward bending ($\cap$ shape) otherwise. Assuming the wing's velocity scales as $V_g \approx \sqrt{2\rho_s g h/\rho}$, and its deceleration is of the same magnitude as the gravitational constant $g$ (see Supplementary document, Figure S.4), we get that, for wings with rigid base and flexible outer edges, the aerodynamic force acting on the flexible edges scales as $\rho WL_fV_g^2/4$ while the gravitational and inertial forces scale as $g\rho_s h_f W  L_f$. The moment-arm for these forces is $L_f/4$, that is half the length of the flexible part on each side. The ratio of aerodynamic to inertial moments is thus given by $M_{\rm aero}/M_{\rm inertia} = h / 2 h_f = 3/2 >1$.   We derive a similar criterion for the wings with rigid edges and flexible base (see Supplementary Document). We get that ${M}_{\textrm{aero}} /{M}_{\textrm{inertia}} <1$ for our experimental conditions. The results of this scaling analysis are summarized in Figure~\ref{fw_fig:parameters_and_patterns}(c) and have two important implications: first, for all non-uniform wings considered here, the ratio $(F_{\textrm{centrifugal}}/F_{\textrm{elastic}})^{1/4}$ is indeed below the critical value  $4.73$ delineating the buckling transition~\cite{TamBushKudrolli2010}. Second, regardless of the flexion ratio $L_r/L$, wings with flexible edges bend upwards ($\cup$ shape) and those with flexible base bend downwards ($\cap$ shape), consistent with our experimental findings.

\begin{figure*}[!t]
   \centering
   \includegraphics[width=0.99\textwidth]{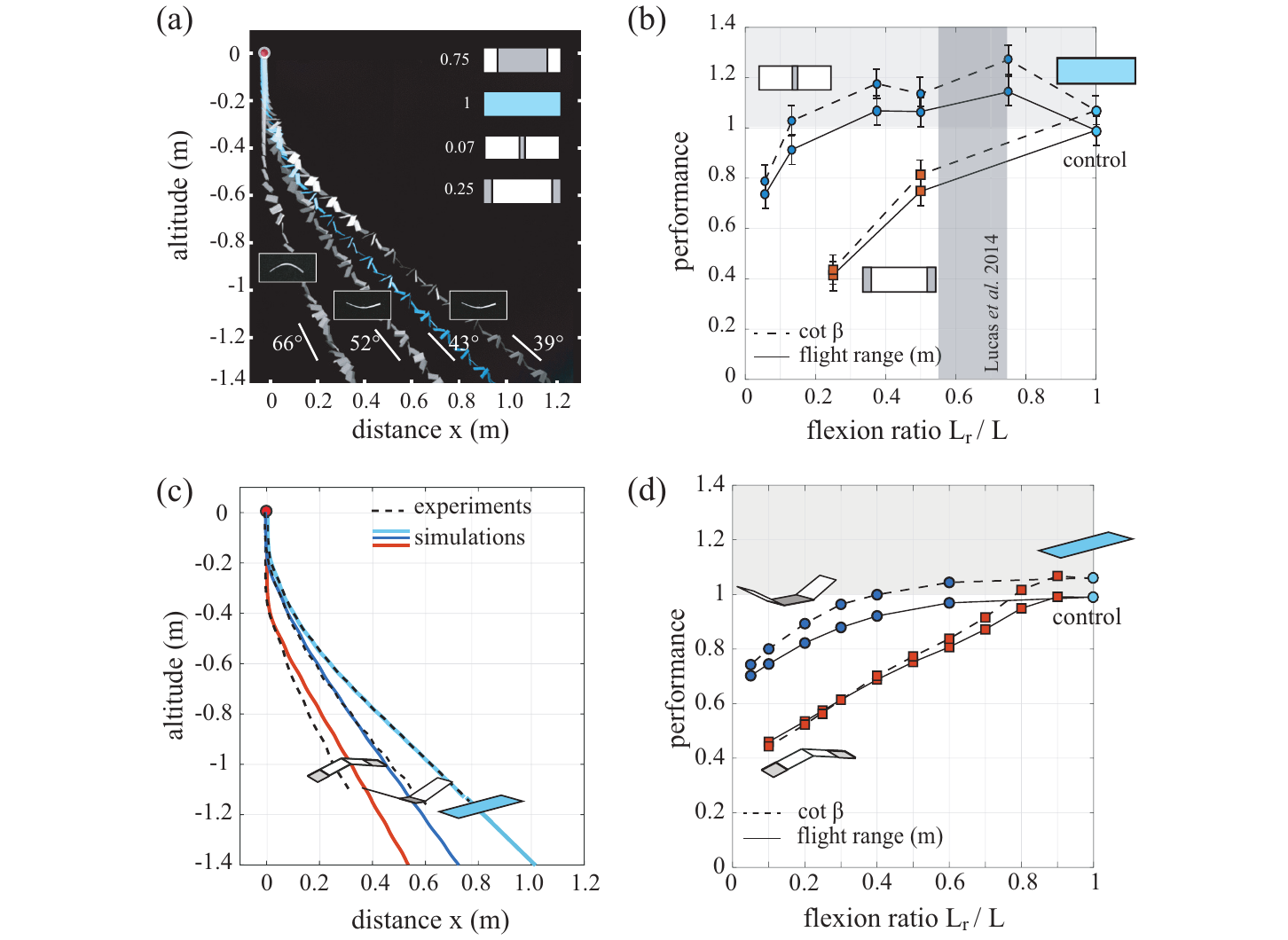} 
   \caption{\footnotesize Effect of spanwise heterogeneous flexibility on tumbling flight. (a) Wings trajectories. (b) Experimental flight range and descent angle measurements show that there is a range of tip flexibility ($0.3 < L_r/L < 1$) associated with an increase of performance, while the performance decreases for higher tip flexibility and any amount of base flexibility. 
(c)   Our quasi-steady model correctly, and quantitatively predict the performance decay for large bending; (d) however, it failed to predict the beneficial window observed in experiments.}
   \label{fw_fig:experiments}
\end{figure*}

\begin{figure}[!t]
   \centering
   \includegraphics[width=0.55\textwidth]{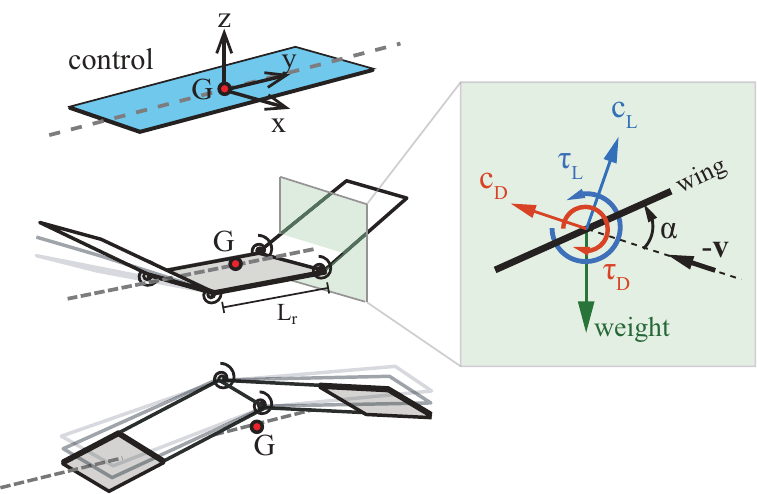} 
   \caption{\footnotesize  
 Quasi-steady aerodynamic model based on blade-element theory. The wings are idealized as an articulated set of rigid panes: two panes for wings with flexible middle and three panes for wings with flexible edges. The spring stiffness at the hinges is adjusted to emulate the wing's bending stiffness. The flapping motion of the panes relative to the wing's rotating frame of reference (attached to its center of mass $G$) modifies the relative wind $-\mathbf{v}$ and the angle of attack $\alpha$ of the moving panes, and thus the wing's aerodynamic behavior.}
   \label{fw_fig:model}
\end{figure}

\paragraph{Flight performance.} We then compared the flight trajectories of the non-uniform wings to that of the control wing; see Figure~\ref{fw_fig:experiments}(a). The control wing exhibited longer flight range (horizontal distance from initial location) and shallower descent angle than the wings with the widest flexible sections at either the wing tips or base. The control wing descended at an angle $\beta  \approx 43\degree$ and reached $x \approx 1$ m whereas $\beta \approx 52\degree$  and $x\approx 0.7$ m for the wing with maximum flexible area at the outer edges and and $\beta \approx 66\degree$ and $x \approx 0.4$ m for the wing with the widest flexible base.
However, the control wing was itself outperformed by a wing with flexible outer edges and $L_r/L = 0.75$; the latter exhibited a shallower descent angle ($\beta \approx 39\degree$ is about $5\degree$ smaller than that of the control wing) and larger flight range  ($x \approx 1.15$ m is about $15\%$ larger than that of the control wing). To put this result in perspective, we compared the flight performance of rigid wings of the same dimensions but different weight. We found that in order to obtain the same improvement in performance with a rigid wing, its weight would have to be reduced by half!  


More generally, the position and spanwise extent of the flexible regions determined flight performance of the wings. Each of the spanwise non-uniform wings described in Table 1 was tested seven times to determine mean values of their descent angles and flight ranges; see Figure~\ref{fw_fig:experiments}(b). We found that wings with a flexible base, irrespective of the flexion ratio $L_r/L$, exhibited diminished flight characteristics (descent angle and flight range) in comparison to the control wing. However, wings with flexible outer edges outperformed the control wing for a wide range of flexion ratio $0.25  \lesssim L_r/L<1$, showing up to 15\% improvement in flight range, with optimal performance for $L_r / L = 0.75$. Interestingly, this ratio agrees with the empirical observations of Lucas \textit{et al.} \cite{LucasCostello2014} on the flexion ratio amid animal propulsors for swimming and flying organisms. This finding supports the hypothesis that improved aerodynamics may be an important selective force in the evolution of animal propulsors.


\paragraph{Comparison to quasi-steady models.} We next compared the experimental results to predictions obtained from a quasi-steady mathematical model.
For the spanwise flexible wings, we idealized the wing as articulated rigid panes (Figure~\ref{fw_fig:model}). The panes flap back and forth about hinges located at the wing's midspan $x=0$ for base-flexible wings, or at $x=\pm L_r/2$ for tip-flexible wings, and endowed with torsional springs of stiffness chosen to emulate the wing's overall flexibility. The flapping motion is entirely slaved to the aerodynamic and inertial forces acting on the panes and is resolved dynamically, together with the wing's translation and rotation. To calculate the aerodynamic forces, we determine the instantaneous local relative wind $-\mathbf{v}$ at each section along the span based on the wing's translation and the motion of segment relative to the wing's center of mass. Underlying the model is the idea that the change in relative wind, and therefore, in angle of attack $\alpha$, modifies the local aerodynamic forces, drag and lift, on each section, and thus the global aerodynamics of the wing. The model and its validation are described in more detailed in the Supplementary Document. 

Figure~\ref{fw_fig:experiments}(c) shows the behavior for two {selected} wings, corresponding to the two most flexible cases for either base-flexible or tip-flexible wings. The model correctly predicted the significant decrease of performance for these two cases.  Figure~\ref{fw_fig:experiments}(d) shows the wing's performance over the entire range of $L_r / L$ considered in our experiments. Notice the gradual deterioration in performance as $L_r / L$ decreases. In contrast to the experimental results, no wing was found to outperform the rigid control $L_r / L=1$. That is to say, the model accurately captured the detrimental effect of flapping for wings with large flexibility, but it failed to reproduce the positive impact of moderate tip flexibility that we observed in experiments. This led us to believe that the interaction between tip-flexibility and the flow is more subtle than what can be captured by a quasi-static, quasi two-dimensional model. To explain the enhancement in performance by tip-flexible wings,  we explored three possible mechanisms for drag reduction.

\paragraph{Candidate mechanisms for drag reduction.}
We explored the effects of planform reduction, streamlining, and wingtip vorticity on drag reduction.
The planform decreases when the wing bends, regardless of where the bending occurs. Since both lift and drag forces are 
proportional to the wing's projected area \cite{Anderson2001}, we expect the the lift-to-drag ratio, and thus the flight performance, to be roughly independent of planform reduction. 
Streamlining, or shape reconfiguration of flexible structures in response to fluid flows, is known to induce a reduction in drag that grows with increasing flow speed~\cite{AlbenShelleyZhang2002,AlbenShelleyZhang2004,DeLangre2008,Gosselin2010}. 
Thus, we expected wings with flexible tips that deform \emph{away from} the incoming flow  to experience reduced drag and wings with flexible middle that deform \emph{towards} the flow to experience increased drag~\cite{Hoerner1965}. Further, we expected drag reduction in tip-flexible wings to monotonically improve with increasing flexibility, and to monotonically worsen in middle-flexible wings. The latter holds true in our experiments, but the change in performance in tip-flexible wings is not monotonic: as the flexion ratio decreases, the performance reaches an optimum then decreases below the control level, suggesting that streamlining, while at play, may not be the right lead to understand the improved performance in tip-flexible wings. 

By shifting focus exclusively to wings with flexible tips, we next explored a novel mechanism that links drag reduction and aerodynamic flows around a fluttering wingtip.  This phenomenon is independent of the wing's autorotation and represents a more general trait of wings possessing flexible tips.


\begin{figure*}
   \centering
   \includegraphics[width=0.99\textwidth]{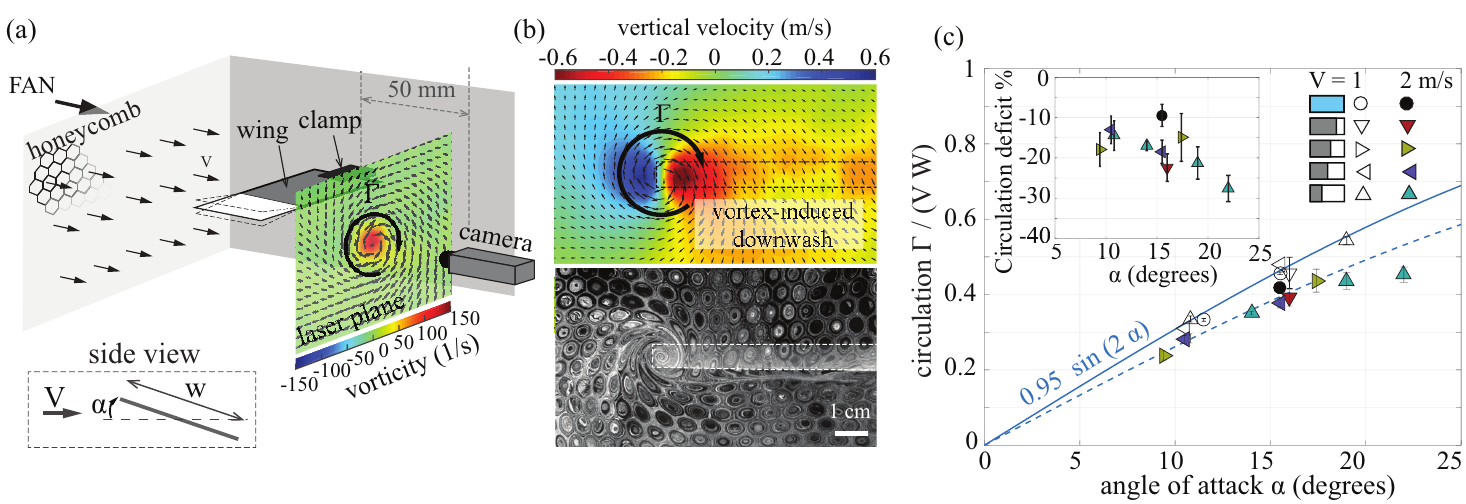} 
   \caption{\footnotesize Flutter-induced drag reduction quantified in wind tunnel experiments: (a) half-wings are set at constant angle of attack $\alpha$ in a low-speed wing tunnel with uniform flow $V$. (B) top panel: D-PIV velocity field highlighting the vortex-induced downwash, that we quantify in terms of the circulation $\Gamma$; bottom panel: qualitative, very low-speed visualization of the wing tip vortex. (C) Vortex circulation in the presence of strong flutter at high flow speed (full symbols) is significantly smaller than for weak flutter at low flow speed (open symbols): we observe a deficit of up to 30\% for high-speed, large flutter cases.}
   \label{fw_fig:wind_tunnel}
\end{figure*}

\paragraph{Flutter-induced drag reduction.} We designed wind tunnel experiments to investigate whether the wings are subject to aerodynamic flutter, and how flutter affects the flow around the wing tip.  Since flutter happens at higher frequencies than the wing's tumbling frequency, we posit that  flutter can be analyzed independently of the wing's rotation. In a custom-made wind tunnel (see Supplementary Document), we placed half-wings at a constant, non-zero, angle of attack $\alpha$ (in the range $9$-$24$\degree) in an oncoming uniform flow at steady speed $V$, as shown in Figure~\ref{fw_fig:wind_tunnel}(a). Using high-speed photography and particle image velocimetry (D-PIV), we quantified the flow in a vertical plane perpendicular to the direction of oncoming flow and located at a distance 26~mm (about 60\% of the chord) downstream from the wing. A representative velocity field is shown in the top panel of Figure~\ref{fw_fig:wind_tunnel}(b) and  the associated vorticity is shown in the inset of Figure~\ref{fw_fig:wind_tunnel}(a) for $V = 2$ m/s. The bottom panel of Figure~\ref{fw_fig:wind_tunnel}(b) was obtained at a lower speed $V\approx 0.1$ m/s only for the purpose of providing a qualitative visualization of the flow downstream of the wing; note the distortion of the upstream honeycomb pattern by the wing.
Figures~\ref{fw_fig:wind_tunnel}(a,b) indicate the presence of a well-formed wingtip vortex, induced by the pressure difference between the lower and upper surface of the wing. Wingtip vortices \cite{Dommasch1967} are known to be detrimental to flight because they induce additional drag due to the ``downwash,'' an additional down-facing component to the velocity field, downstream from the wing's trailing edge \cite{Green1995,BirchDickinson2001}. Downwash is clearly visible in Figure~\ref{fw_fig:wind_tunnel}(a).
The downwash-induced drag increases with angle of attack and decreases with aspect ratio, and it contributes significantly to the total drag acting on the wing \cite{RaspaRamananarivo2014}. 

We tested experimentally whether flutter due to wing tip flexibility decreases downwash and therefore downwash-induced drag. To this end, we considered that the circulation intensity $\Gamma$ of the wingtip vortex is a reasonable integral measurement of the downwash: larger $\Gamma$ indicates stronger downwash and larger downwash-induced drag. To quantify the influence of tip flutter on downwash, we measured both the flutter amplitude and the circulation $\Gamma$ at two flow speeds $V=1$ m/s and $V = 2$ m/s using the same set of wings employed in the tumbling flight experiments and at various angles of attack. At  $V=1$ m/s, we observed little or no deflection for all wing tips. At $V = 2.0$ m/s, we obtained significantly higher amplitudes of tip flutter, in the range $2-30$ mm (aerodynamic forces scale as $V^2$). Note that the translational speed observed in the tumbling flight experiments is greater than $V = 1.5$ m/s and suggest the presence of moderate flutter. 
For each wing, angle of attack, and flow speed, we calculated $\Gamma$ using a time-averaged version of the measured flow field with time-average window of $0.5$~s, corresponding to about six flutter periods of the most flexible wing. 
For this range of oncoming flow speeds and associated Reynolds numbers, the two-fold increase in $V$ has little to no effect on the flow topology, as suggested by \cite{Spedding1992,Schmitz1941}, and 
 $\Gamma$  scales as $V W$ for a rigid wing~\cite{Dommasch1967}. We thus used $\Gamma / V W$ as a non-dimensional measure of the circulation. Figure~\ref{fw_fig:wind_tunnel}(c) shows that data for $V = 1.0$ m/s (weak flutter) collapse well on a master curve well fitted by $\Gamma / (V W) = 0.88 \sin(\alpha)$ for all angles of attack. However, data for $V = 2.0$ m/s (strong flutter) are 10\% to 30\% lower than the master curve (15\% deficit curve is shown for reference). These results suggest that tip flutter induces a substantial decrease of wing tip vortex downwash, which decreases the amount of induced drag and therefore improves the wing's performance.


\paragraph{Chordwise-heterogeneous wings.} 
To complete our analysis of the behavior of non-uniform wings, we investigated the effect of changing the mass distribution chordwise; see right column of Table~\ref{fw_table:prop_wings}. 
This chordwise non-uniformity did not induce wing deformations under the aerodynamic and gravitational loads considered here. However, the dimensionless moment of inertia $I$ varied by nearly one order of magnitude. We thus anticipated that it would have a substantial effect on the flight behavior.

\begin{figure*}
   \centering
   \includegraphics[width=0.99\textwidth]{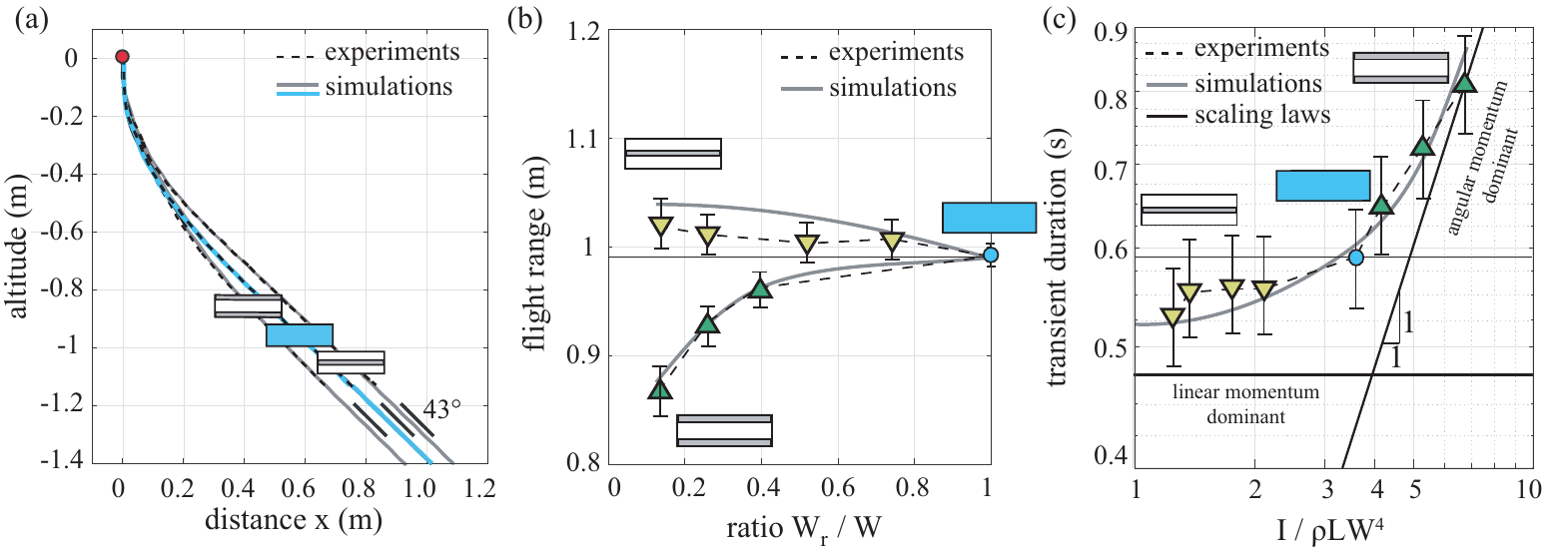} 
   \caption{\footnotesize Effect of chordwise mass heterogeneity on tumbling flight. (a-b) Wings trajectories and flight range measurements show that larger moment of inertia around the tumbling axis lead to smaller flight range. (c) The performance change is due to a longer transient resulting from the increase of moment of inertia, in excellent agreement with simulations (thick grey line) and scaling arguments (black line).}
   \label{fw_fig:chordwise}
\end{figure*}

In Figure~\ref{fw_fig:chordwise}(a), we compared the behavior of two representative wings to the control wing. The wing's layout had a clear effect on the flight range: wings with the heavier portions located away from the center of rotation (larger moment of inertia,\,{\tikz\draw[black,fill=GreenYellow,thick] (4.55ex,1.15ex) -- (5.85ex,1.15ex) -- (5.2ex,0ex) -- cycle;}) showed smaller flight range, while wings that are heavier closer to the center (smaller moment of inertia, {\tikz\draw[black,fill=Green,thick] (4.35ex,0ex) -- (5.65ex,0ex) -- (5ex,1.15ex) -- cycle;}) showed improved performance. This is supported in Figure~\ref{fw_fig:chordwise}(b) by the monotonic trend observed for each family of wings. Interestingly, the terminal descent angles of all wings is identical (43$\pm 1 \degree$), which suggests that the wing's layout primarily affected the duration of the transient regime. 

To corroborate this assumption, we measured the duration of the transient regime, defined as the time needed for the wing to reach a phase-average value within 2\degree {\tiny~}of the terminal descent angle. The results are shown in Figure~\ref{fw_fig:chordwise}(c) as a function of the wings' moment of inertia. Clearly, wings with heavier outer portions (larger moment of inertia) exhibited significantly longer transient periods than the control, while wings with heavier middle portion (lower moment of inertia) showed shorter transience. We then used our quasi-steady model to provide a prediction of the trajectory, flight range and transient duration for all tested wings (thick grey lines in Figure~\ref{fw_fig:chordwise}(a-c). We obtained excellent agreement with the experiments by only changing the moment of inertia to match that of the actual wings. Taken together, these results demonstrate that the wing's flight range depends on the duration of the transient regime, which in turn is directly related to the wing's moment of inertia.

A peculiar characteristic of the flight characteristics is the sharp decrease in flight range for wings with thicker edges, particularly evident for $W_r / W < 0.4$ (Figure~\ref{fw_fig:chordwise}b). This is correlated with a sharp increase in the duration of the transient regime (Figure~\ref{fw_fig:chordwise}c). In contrast, we only observed gradual changes for wings with thick middle, which suggests that the detrimental effect of moment of inertia progressively vanishes as the latter decreases.

This transition happens when the process governing the transient regime changes: for small moment of inertia, linear momentum is limiting, while for large moment of inertia, angular momentum is limiting. A simple scaling argument (see Supplementary Document for further details) shows that when linear momentum is dominant, the duration  of transient regime is independent of the non-dimensional moment of inertia $I/\rho L W^4$, whereas when angular momentum is dominant, the transient time scales linearly with $I/\rho L W^4$. A transition  is expected when the two time scales are equal, which corresponds to a non-dimensional moment of inertia $(I/\rho L W^4) \approx (\pi \rho_s h/ 8\rho W) \textrm{St}^{-1} \approx 4.0$. This prediction is in good agreement with the experimental data in Figure~\ref{fw_fig:chordwise}(c).


\section*{Conclusion}

We investigated the flight characteristics (flight range and descent angle) of wings of the same mass $1.3$ g and planform $120 \times 42$ mm, but non-uniform spanwise and chordwise mass distribution. We analyzed the behavior of these wings by means of  flight experiments, quasi-steady mathematical models, scaling analysis, and wind-tunnel experiments. 

The non-uniform mass distribution in the spanwise direction caused heterogeneous changes in the wing's flexibility. Wings with flexible middle experienced diminished flight compared to the control wing, whereas wings with flexible edges exhibited an improvement in flight performance for a range of flexion ratios. Interestingly, optimal performance was observed at flexion ratios consistent with the universal two-third bending rule reported in~\cite{LucasCostello2014} for a broad spectrum of animal propulsive appendages. Closer investigation of the mechanisms leading to this improvement in performance points to flutter-induced drag reduction caused by a reduction in wing-tip vorticity around these flexible appendages. This phenomenon is independent of the wings' auto-rotation; it is expected to hold as a universal mechanism for drag reduction around all flexible appendages.

To complete this study, we examined the effect of chordwise mass heterogeneity. The non-uniform mass distribution in the chordwise direction led to first-order changes in the wing's moment of inertia, with larger moment of inertia inducing longer transience before settling into periodic tumbling, thus leading to shorter flight range.

Taken together, these findings have important implications on the optimal design of tumbling wings assigned the task of transporting a given load (mass) the farthest distance possible. To improve its flight range, the wing would have to distribute its mass non-uniformly such that the wing's tips are flexible and its moment of inertia is minimal. That is to say, the wing would have to concentrate most of the mass at the center, exactly the design observed in several tumbling seedpods such as in the seedpods of \textit{Ailanthus altissima} \cite{Kowarik2007}.

To conclude, we note that while we only showed experimental data for one planform and wing thickness, the results presented here are valid for a wide range of parameters. Specifically, we repeated the experiments for wings of various mass and planforms and found identical trends: an improvement in flight performance for wings with moderate tip flexibility.

Since flexibility plays a key role in bio-inspired soft robotics \cite{Kim2013, Tolley2014,Rich2018} and tissue-engineered systems \cite{Nawroth2012,ParkParker2016}, we expect our methods to provide a framework from which to understand and design flexible appendages that flap \textit{actively} to generate propulsive forces.  Our findings suggest that it is beneficial to engineer propulsors with flexible outer edges that do not directly contribute to active force generation, but that indirectly, through fluid interactions, enhance the overall performance of the propulsor~\cite{Colin2012}. 

%
%
%
%

\appendix




\section{Scaling analysis: Spanwise non-uniform wings} \label{section:scaling_argument}

We present here a general argument of bending direction, valid for wings made of $n+1$ layers of paper ($n=2$ in our work), each of thickness $h_f$. We derive the calculations of bending moments for both types of spanwise non-uniform wings: flexible middle and flexible tips. The wing is subject, on average, to a relative wind equal to  the terminal velocity $V_g \approx \sqrt{2\rho_s g (n+1)h_f/\rho}$, while its velocity changes with time with a typical acceleration/deceleration that is approximately equal to the free-fall acceleration $g$, in agreement with experimental data (see Figure~\ref{fw_fig:fit_control}b). Maximum bending occurs when the wing is horizontal. At this instant, the aerodynamic force acts directly upwards and the gravitational and inertial forces directly downwards.

\begin{figure}[!h]
\center
\includegraphics[width=0.8\textwidth]{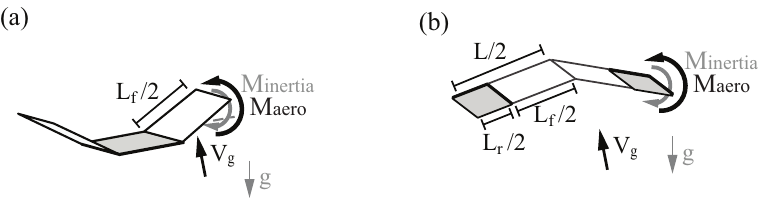}
\caption{\footnotesize (a) Details of aerodynamic and inertial loads on wings with (a) rigid base and (b) flexible base. The downward or upward bending depends on the ratio of aerodynamic-to-inertial loads.}
\label{fig:scaling_arguments}
\end{figure}

For wings with rigid base and flexible edges, see Figure~\ref{fig:scaling_arguments}(a), the aerodynamic force acting on (one) flexible edge of length $L_f/2$ scales as $\frac{1}{2} \rho W (L_f/2) V_g^2$ and has a moment arm $L_f/4$, which creates the following aerodynamic moment:
\begin{equation}
M_{aero} = \frac{1}{2} \rho W \frac{L_f^2}{8} V_g^2 = \rho_s g W \frac{L_f^2}{8} (n+1) h_f
\end{equation}
Following a similar argument, the inertial moments, from weight and wing's inertia, is:
\begin{equation}
M_{inertia} = \rho_s (a+g) W \frac{L_f^2}{8}  h_f 
\end{equation}
The ratio of aerodynamic to inertial moments is:
\begin{equation}
\frac{M_{aero}}{M_{inertia}} = \frac{n+1}{1+a/g} \approx \frac{3}{2} > 1
\end{equation}
resulting in upward deflection.

For wings with flexible base and rigid edges, see Figure~\ref{fig:scaling_arguments}(b), we write the balance on the full half-wing of length $L/ 2$. The aerodynamic moment about the wing's middle is:
\begin{equation}
M_{aero} = \frac{1}{2} \rho W \frac{L}{8} V_g^2 = \rho_s g W \frac{L^2}{8} (n+1) h_f
\end{equation}
The inertial moment consists of two contributions: that of the base layer, spanning the entire half-wing (moment arm $L/4$) , and that, of the rigid sides made of the n folded layers (moment arm $L_f/2 + L_r/4$). Combining the two  contributions:
\begin{equation}
M_{inertia} = \rho_s (a+g) \frac{L^2}{8} W h_f \left[ 1 + n \left(2 - \frac{L_r}{L} \right) \right]
\end{equation}
The ratio of moments is:
\begin{equation}
\frac{M_{aero}}{M_{inertia}} = \frac{n+1}{(1+a/g) \left[ 1+n(2-L_r/L) \right] } \approx \frac{3}{2 ( 1+2(2-L_r/L) )} < 1 \quad \text{for} \ L_r / L < 7/4 
\end{equation}
Downwards deflection is guaranteed for all attainable values of flexion ratio ($0 < L_r / L \leq 1$). 
This scaling analysis predicts the \emph{direction} of bending but does not predict the bending amplitude.





\begin{figure}[!t]
   \centering
   \includegraphics[width=0.99\textwidth]{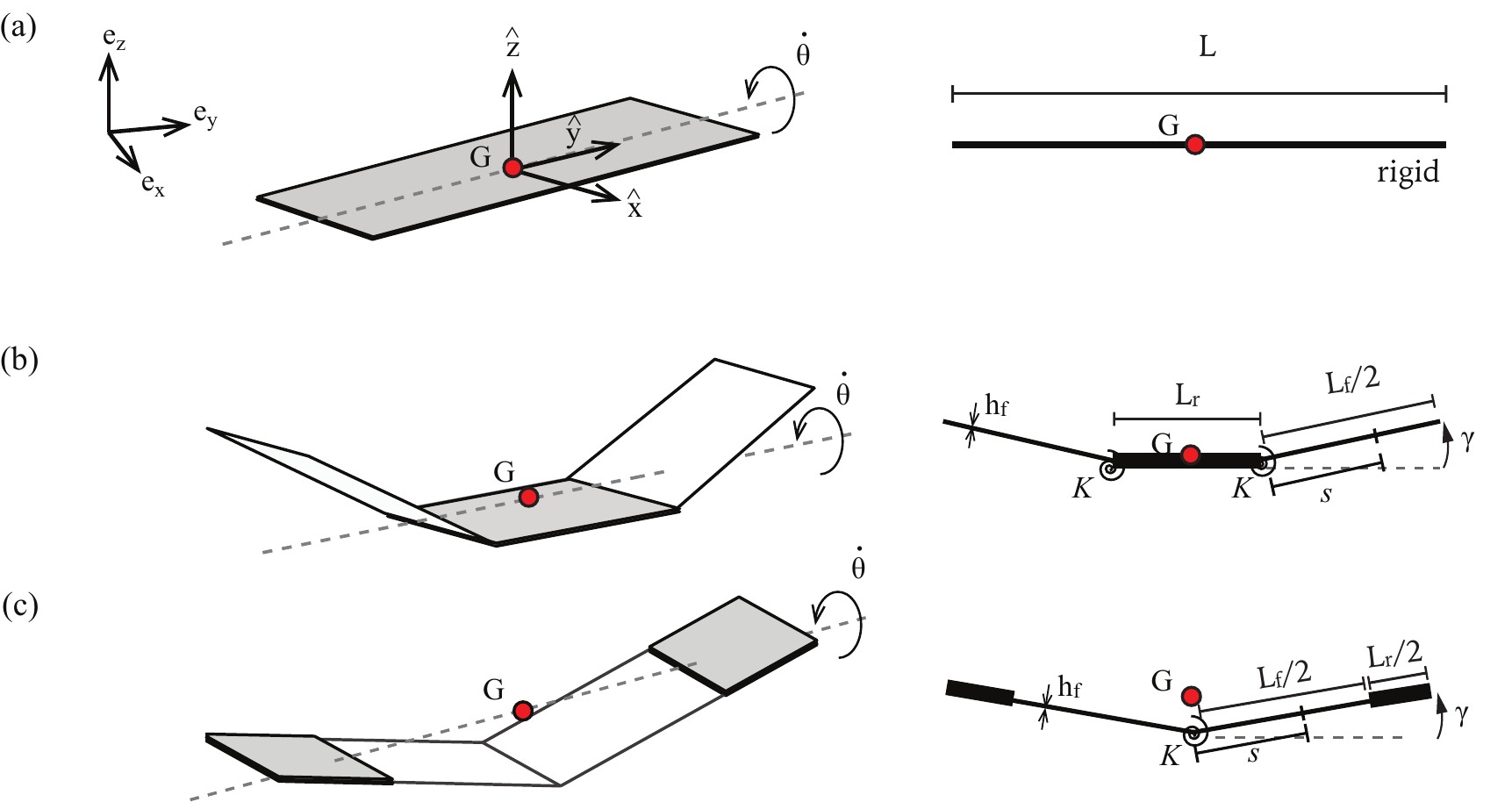} 
   \caption{Model and wing idealization for the blade-element theory for (a) the rigid wing (control), (b) wings with flexible tips, and (c) wings with flexible base.}
   \label{fw_fig:BEM_model}
\end{figure}


\section{Control wing: quasi-steady force model} \label{section:model}
\label{sec:control}

\paragraph{Wing kinematics.} The wing tumbles in a vertical $(x,z)$-plane, where the $x$-axis defines the horizontal flight range and the $z$-axis points vertically upward. It is convenient to introduce an inertial frame of reference $({\mathbf{e}_x},{\mathbf{e}_y},{\mathbf{e}_z})$ and an orthonormal auto-rotating frame $(\hat{\mathbf{x}},\hat{\mathbf{y}},\hat{\mathbf{z}})$ attached to the center of the wing, where $\hat{\mathbf{x}}$ and $\hat{\mathbf{y}}$ are unit vectors along the chordwise and spanwise directions.

The center of mass of the wing can be described by the position vector $\mathbf{r} = x \mathbf{e}_x + z \mathbf{e}_z$, its translational velocity is $\mathbf{v} = \dot{x} \mathbf{e}_x + \dot{z} \mathbf{e}_z$, which can be written in the co-rotational frame as $\mathbf{v} = v_{x} \hat{\mathbf{x}} + v_{z} \hat{\mathbf{z}}$. 
Equivalently, we can rewrite $\mathbf{v}= v (\cos\alpha \hat{\mathbf{x}} +  \sin\alpha \hat{\mathbf{z}})$, where $\alpha$ is the angle of attack $\alpha$ between the velocity vector $\mathbf{v}$ and the $\hat{\mathbf{x}}$-direction and $v = \| \mathbf{v} \|$ is the translational speed of the wing. The wing rotates at an angular velocity $\dot{\theta}$ about the $\hat{\mathbf{y}}$-direction. Here and thereafter, we use the dot notation $\dot{()}$ to denote differentiation with respect to time. 

\paragraph{Balance laws.}
We write  the balance of linear and angular momenta on the wing using a quasi-steady aerodynamic force and moment model (see, for example, \cite{AndersenWang2005})
\begin{equation}
\begin{split}
m \dot{\mathbf{v}} &=  m \mathbf{g} + \mathbf{F}_L +  \mathbf{F}_R +  \mathbf{F}_D,   \\
I \ddot{\theta}  &=  T_L - T_D .
\label{eq:linear}
  \end{split}
\end{equation}

\begin{figure}[!h]
\center
\includegraphics[width=0.99\textwidth]{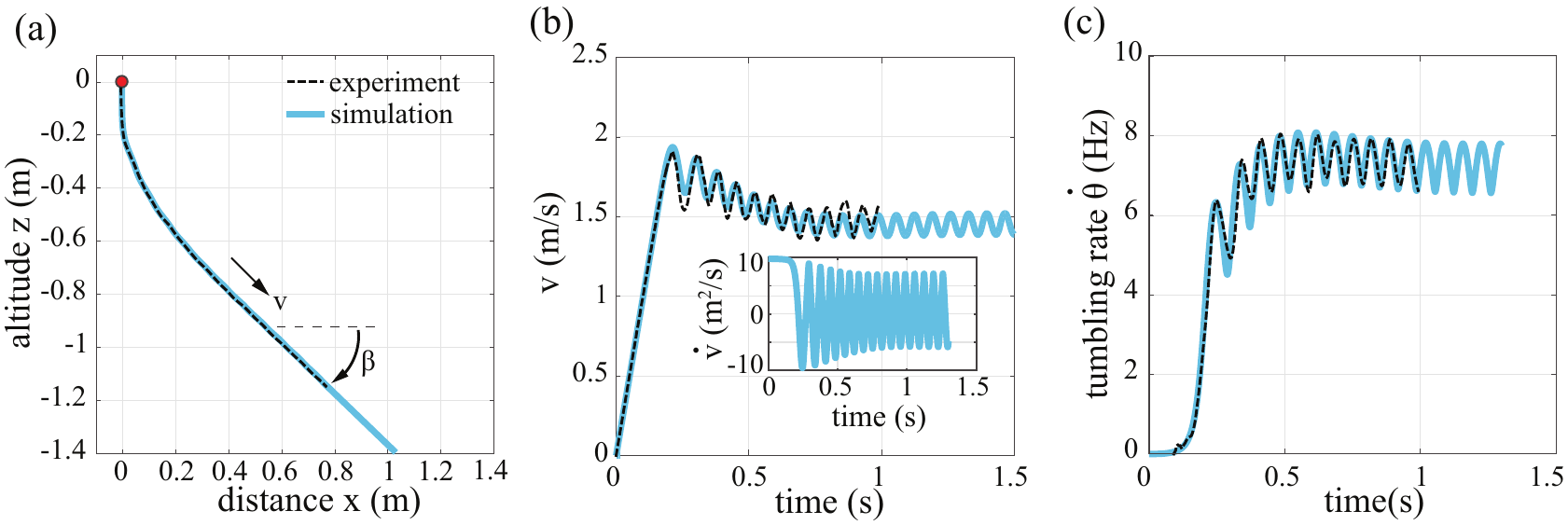}
\caption{(a) Trajectory, (b) velocity and (c) tumbling rate of the control rigid wing, predicted (thick grey line) and experimental (dotted black line). The inset in (b) shows the time-dependent linear acceleration: the wing experiences maximum accelerations and decelerations of the order of $ g=9.81$ m/s$^2$ throughout its flight. The prediction faithfully predicts the characteristics of the permanent regime as well as fine features of the transient regime.}
\label{fw_fig:fit_control}
\end{figure}

Here, $m$ is the total mass of the wing, $I$ its moment of inertia about the spanwise $\hat{\mathbf{y}}$-direction, and $\mathbf{g}=-g\mathbf{e}_z$ is the gravitational acceleration, pointing vertically downwards.
The aerodynamic forces acting on the wing consist of a drag force $\mathbf{F}_D$ parallel to $-\mathbf{v}$ and translational and rotational lift forces $\mathbf{F}_L$ and  $\mathbf{F}_R$  perpendicular to $-\mathbf{v}$; see the bottom panel of Figure~\ref{fw_fig:fit_control}(a). Specifically, we have
\begin{equation}
\mathbf{F}_D = - {F}_D \dfrac{\mathbf{v}}{\|\mathbf{v}\|}, \qquad \mathbf{F}_L = - {F}_L \dfrac{\mathbf{v}^\perp}{\|\mathbf{v}\|}, \qquad  \qquad \mathbf{F}_R = - {F}_R \dfrac{\mathbf{v}^\perp}{\|\mathbf{v}\|},
\end{equation}
where $\mathbf{v}^\perp= v(-\sin\alpha \hat{\mathbf{x}} + \cos\alpha\hat{\mathbf{z}})$ is perpendicular to $\mathbf{v}$.
The aerodynamic moments act about the $\hat{\mathbf{y}}$-direction and can be decomposed into a moment $T_L$ induced by the lift forces and a dissipative moment $T_D$.

\begin{figure}[!h]
\center
\includegraphics[width=0.65\textwidth]{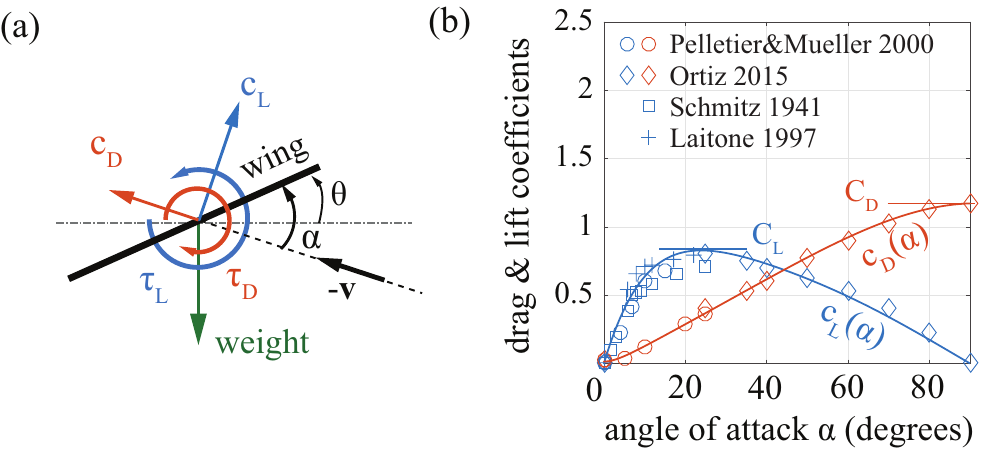}
\caption{(a) Forces and torques acting on the wings in our quasi-steady model. (b) The drag and lift coefficients used in our quasi-steady model (solid lines) are fitted to existing experimental data for fixed plates at comparable Reynolds number  \cite{PelletierMueller2000,Ortiz2015,Schmitz1941,Laitone1997}. }
\label{fw_fig:CD_and_CL}
\end{figure}

\paragraph{Aerodynamic force model.} The magnitude of the drag force $\mathbf{F}_D$, translational lift $\mathbf{F}_L$, and the rotational lift $\mathbf{F}_R$ are given by
\begin{eqnarray}
F_L = \frac{1}{2} c_L(\alpha) \rho L W v^2, \qquad F_D =  \frac{1}{2} c_D(\alpha) \rho L W v^2, \qquad
F_R = \rho  \Gamma_{\! R} v .
\label{eq:force}
\end{eqnarray}
Here, the drag and lift coefficients  $c_D(\alpha)$ and $c_L(\alpha)$ are not known apriori. To obtain their dependence on $\alpha$, we 
analyze and extract the functional form  of $c_D$ and $c_L$ from data of  drag and lift coefficients found in the literature for comparable Reynolds numbers \cite{PelletierMueller2000,Ortiz2015,Schmitz1941,Laitone1997}, as shown in Figure~\ref{fw_fig:CD_and_CL}(b), but we allow the pre-factor scaling each function to be an adjustable parameter. Specifically, we let  $C_L$ and $C_D$, corresponding to the maximum lift and maximum drag be these adjustable constants, as shown in Figure~\ref{fw_fig:CD_and_CL}(a). 

The wing's rotation generates a circulation $\Gamma_{\! R}$ proportional to the tip velocity $W\dot{\theta}$ and the wing's chord $W$~\cite{AndersenWang2005, PesaventoWang2004}; Specifically, one has $\Gamma_R =  \frac{1}{2} C_{\!R} W^2 \dot{\theta}$, where $C_R$ takes a constant value, independent of the angle of attack $\alpha$. According to the Kutta-Joukowski theorem, the magnitude of the associated rotational lift is $F_R = \rho  \Gamma_{\!R} L v$ as noted in~\eqref{eq:force}.

\paragraph{Torque model.} The rotational torque results from the translational lift force along the span
\begin{equation}
T_L  = e  W F_L \dfrac{\alpha}{|\alpha|}.
\end{equation}
Here, $e$ is a parameter representing the lever arm of the lift force, expressed as a fraction of the chord W ~\cite{AndersenWang2005}. This torque can be driving or hindering the wing rotation depending on the sign of $\alpha$. The dissipative torque takes the form 
\begin{equation}
T_D = \mu \rho L W^4  \dot{\theta} |\dot{\theta}|,
\end{equation}
where $\mu$ is an adjustable parameter \cite{AndersenWang2005,TamBushKudrolli2010};  in the quasi-permanent tumbling regime, $T_D$ and $T_L$ balance each other on average every half-revolution.

\paragraph{Adjustable parameters}

To close the model \eqref{eq:linear}, we need to estimate the non-dimensional parameters $C_D$, $C_L$, $C_R$, $e$, and $\mu$. We used the values given in~\cite{AndersenWang2005} as initial guess.
Then, we adjusted these parameters in order to match the characteristics parameters of the descent of a rigid wing, namely the descent angle, time-dependent velocity $\mathbf{v}$ and instantaneous rotational velocity $\dot{\theta}$. We get $C_{L} = 1.12$, $C_{R} = 0.94$, $C_{D} = 2.07$, $e = 0.18$ and $\mu = 0.024$. These values remain constant throughout this document.
Finally, to reconstruct the full dynamics from~\eqref{eq:linear}, we need to specify initial conditions for the position, linear velocity, pitch orientation and angular velocity. Here, we have $(x,z)$=(0,0), $\mathbf{v}(0)=0$, $\dot{\theta}(0)=0$ and $\theta(0) = -90\degree + \varepsilon$, where $\varepsilon$ is fixed at $0.7 \degree$ for all tests.

\paragraph{Model validation}
Figure \ref{fw_fig:fit_control}(a-c) shows that the time-integration of our system of equations \eqref{eq:linear} with these parameters value is in excellent agreement with experimental data for the control (rigid) wing: trajectory, velocity and angular velocity all match quantitatively.

\begin{table}[!t]
\begin{center}
\begin{tabular}{l*{7}{c}r}
Model             & $C_L$& $C_R$ & $C_D$ & e & $\mu$  & K  & $k_s$ \\
\hline
\hline
Rigid   & 1.12 & 0.94 & 2.07 & 0.18 & $0.024$ &   - & - \\
\hline
Flexible tips  & 1.12 & 0.94 & 2.07 & 0.18 & $0.024$  & $8 E J / L_f$    & $ 4.5 \times 10^{-5} L_f/2$  \\
Flexible middle  & 1.12 & 0.94 & 2.07 & 0.18 & $0.024$ & $4 E J / L_f$    & $4.5 \times 10^{-5} L_f$  \\
\end{tabular}
\caption{Model parameters}
\label{table:model_param}
\end{center}
\end{table}

\section{Spanwise non-uniform wings: blade-element models} \label{section:model_nonuniform}

\paragraph{Wings with flexible outer edges.} The wing model consists of three articulated rigid panels (or `blades'): two identical outer panels of length ${L_f}/2$ and thickness $h_f$ and a middle panel of length $L_r$ and thickness $h_r$, all of width $W$. Each outer panel is connected via a hinge joint along one edge to the middle panel and free to rotate about that edge as shown in Figure~\ref{fw_fig:BEM_model}(b).  To emulate the wing's elasticity, the hinge joints are equipped with torsional springs of constant stiffness $K$. The stiffness $K$ depends on the wing's layout, geometry, and material properties as detailed in section~\ref{sec:stifness} and summarized in Table \ref{table:model_param}.
The outer panels rotate symmetrically relative to the middle; their relative rotation angle is denoted by $\gamma$.
The wing deflection is related to $\gamma$; the maximum deflection $\zeta$ is given by $\zeta = ({L_f}/{2})\sin{\gamma}$. 
. 

As in the case of the control wing, an auto-rotating frame $(\hat{\mathbf{x}},\hat{\mathbf{y}},\hat{\mathbf{z}})$ is attached at the center of the middle panel. The vector $\mathbf{v}$ denotes the translational velocity of the center of the middle panel, whereas $\dot{\theta}$ denotes the angular rotation of the middle panel about $\hat{\mathbf{y}}$. The angular velocity vector of the outer two panels is given by  $\dot{\theta} \hat{\mathbf{y}} +\dot{\gamma}\hat{\mathbf{x}}$. 
It is convenient for computing the aerodynamic forces and moments acting on the wing to introduce the velocity vector $\mathbf{v}_r$ such that $\mathbf{v}_r = \mathbf{v}$ on the middle panel and $\mathbf{v}_r = \mathbf{v} + s\dot{\gamma}(\cos \gamma \hat{\mathbf{z}} \pm \sin\gamma\hat{\mathbf{y}} )$ along the outer panels. Here, $s$ is a spanwise distance measure from the hinge joint.

The motion of the deformable wing is thus described by four degrees of freedom: two associated with the translational motion of the center of the middle panel, one associated with the tumbling rotation, and one associated with the deflection of the outer panels.

Equations~\eqref{eq:linear} provide three scalar equations of motions for the translational and tumbling motion of the wing. To compute the aerodynamic forces in~\eqref{eq:linear}, we assume that each section of the wing is subject to local drag $\mathbf{f}_D$ and translational lift $\mathbf{f}_L$ densities (forces per unit length), that vary along the span of the wing. Specifically, we have
\begin{equation}
\mathbf{f}_L = - f_L \dfrac{\mathbf{v}_r^\perp}{\| \mathbf{v}_r\|}, \qquad 
\mathbf{f}_D = - f_D \dfrac{\mathbf{v}_r}{\| \mathbf{v}_r\|},
\end{equation}
where
\begin{equation}
f_L = \frac{1}{2}  c_{L}(\alpha_r) \rho W v_r^2, \qquad 
f_D = \frac{1}{2} c_{D} (\alpha_r) \rho W v_r^2.
\end{equation}
 Here, the angle of attack $\alpha_r$ is defined as the angle between $\mathbf{v}_r$ and the wing's pitch ($\hat{\mathbf{x}}$-axis), and it is a function of the section of wing under consideration, as shown in Figure~\ref{fw_fig:BEM_model}(b). 
The total drag $\mathbf{F}_D$ and translational lift $\mathbf{F}_L$ are obtained by integrating these  force densities along the span of the wing ($d\ell$ is an infinitesimal distance along the span of the wing)
\begin{eqnarray}
\mathbf{F}_L = \int_0^L \mathbf{f}_L d\ell, \qquad \mathbf{F}_D = \int_0^L \mathbf{f}_D d\ell.
\end{eqnarray}
%
Similarly, the rotational torque $T_L$ is calculated as the integral of a torque density (applied by the translational lift force) along the span:
\begin{equation}
T_L =  \int_{0}^{L} e  W f_L \alpha_r / |\alpha_r| d\ell.
\end{equation}
The expressions for the remaining forces and moments in~\eqref{eq:linear} are the same as described previously. 

In addition to~\eqref{eq:linear}, we need to formulate an equation of motion for $\gamma$. 
To this end, we write the balance of angular momentum on one of the outer panels. We have
\begin{equation}
I_x \frac{d^2\gamma}{dt^2} =  \mathcal{M}_{\textrm{aero}} + \mathcal{M}_{\textrm{inertia}} + \mathcal{M}_{\textrm{centrif}} +  \mathcal{M}_{\textrm{elastic}} + \mathcal{M}_{\textrm{dissip}} \label{EOM_bending}.
\end{equation}
Here, $I_x= \rho_s h_f W (L_f/2)^3 / 3$ is the moment of inertia of the outer panel about an axis along the hinge joint connecting it to the middle segment.  The aerodynamic moment $\mathcal{M}_{\textrm{aero}}$ is given by
\begin{equation}
\mathcal{M}_{\textrm{aero}} =  \int_{s=0}^{L_f/2} \left[s(\cos\gamma\hat{\mathbf{y}} + \sin\gamma \hat{\mathbf{z}})\times (\mathbf{f}_L(s) + \mathbf{f}_D(s))\right] \cdot \hat{\mathbf{x}}ds,
\end{equation}
where $s=0$ corresponds to the location of the wing's hinge joint. 

The moment $\mathcal{M}_{\textrm{inertia}}$ induced by the inertia (weight and acceleration) of the outer panel is given by
\begin{equation}
\mathcal{M}_{\textrm{inertia}} = -  \dfrac{1}{2}( g \cos{\theta} + \dot{\mathbf{v}}_r . \mathbf{e}_z )  \rho_s h_f W \left(\dfrac{L_f}{2}\right)^2  
\end{equation}
 The centrifugal moment $\mathcal{M}_{\textrm{centrif}}$ induced by the wing's rotation can be approximated by
\begin{equation}
\mathcal{M}_{\textrm{centrif}} \simeq \dfrac{1}{3}\rho_s h_f W \dot{\theta}^2 \sin{\gamma} \left( \dfrac{L_f}{2}\right)^3 
\end{equation}
The moment $\mathcal{M}_{\textrm{elastic}}= -K \gamma$ is  the restoring elastic torque imposed by the torsional spring, and the moment $\mathcal{M}_{\textrm{dissip}} = - k_s \dot{\gamma}$ accounts for all dissipation sources other than aerodynamic. 

\paragraph{Wings with flexible middle.} The model consists of two rigid panels of non-uniform thickness connected along one edge in the middle and free to rotate relative to each other about that edge, as shown in Figure~\ref{fw_fig:BEM_model}(c).  Similar to the wings with flexible outer edges, we use a blade element model to compute the aerodynamic forces and moments applied on the wing.
For wings with flexible middle:
\begin{eqnarray} 
\mathcal{M}_{\textrm{aero}} &=& \int_{s=0}^{L/2} s f_\perp(s) ds \\
\mathcal{M}_{\textrm{inertia}} &=& - ( g \cos {\theta} + \dot{\mathbf{v}} . \mathbf{z} ) \rho_s W h_f [  L_f^2 + (2 / \lambda + 1) (L^2-L_f^2) ]/8 \\
\mathcal{M}_{\textrm{centrif}} &\simeq& \rho_s h_f W L^3 \dot{\theta}^2  \sin{\gamma} \left[ \lambda (5 \lambda -3) + 1 \right] / 24
\end{eqnarray}
where $\lambda = L_r / L$. In this case, $s=0$ is at the wing's middle and the half wing's moment of inertia is $I_x = \rho_s W  h_f / 24 \left[ (2/ \lambda + 1) L^3  - (2/ \lambda) L_f^3 \right ]$.

\begin{figure}[htb]
\center
\includegraphics[width=0.89\textwidth]{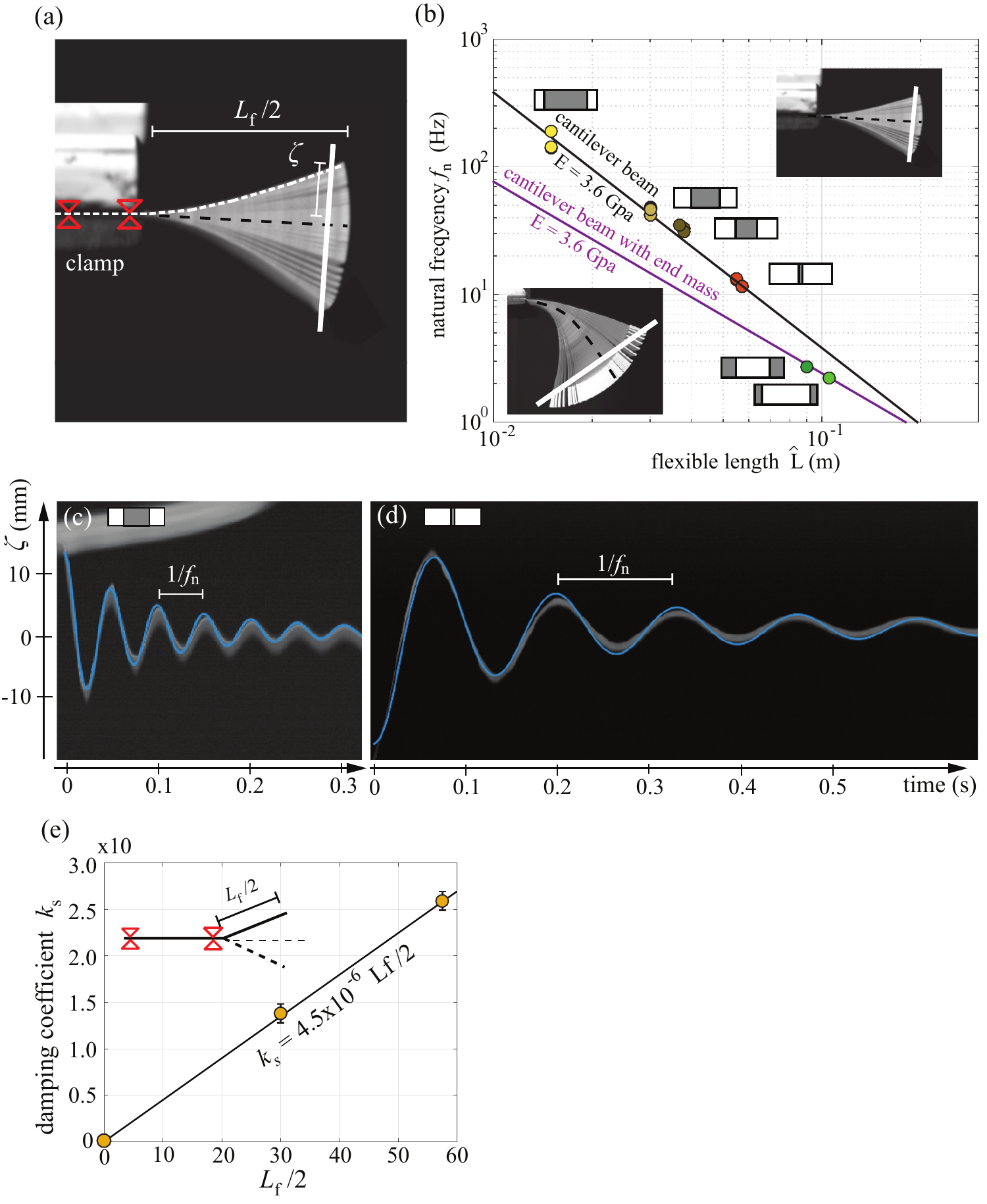}
\caption{Mechanical characterisation of the flexible wings. (a-d) Experimental determination of the natural frequency $f_n$, leading to the Young's modulus $E$ using tip deflection predictions of $f_n$ using beam theory. (c-d) are space-time diagram of the wing tip relaxation dynamics for  (c) $L_\textrm{f}/2$ = 30 mm and (d) $L_\textrm{f}/2$ = 57.5 mm and an initial deflection of $\zeta_{\textrm{t=0}} \simeq 15$ mm. Extraction was performed in a direction normal to the equilibrium position, shown in (a-b). (e) Estimation of the friction-related damping coefficient $k_s$ in the one-hinge model. Predicted tip dynamics is shown in blue in (c-d).} 
\label{fw_fig:oscillograms}
\end{figure}

\section{Measuring wing's stiffness and damping parameters} \label{sec:stifness}

The torsional spring constant $K$ and the damping coefficient $k_s$ are adjusted to fit that of a real wing. We calculate an equivalent $K$ so that the tip deflection $\zeta$ is identical for both the simplified and the continuous wing. We define the quantity $\hat{L}$ as the length of the uninterrupted flexible portion: $\hat{L} = L_f / 2$ for wings with flexible tips, and $\hat{L} = L_f$ for wings with flexible middle. For a cantilever beam with uniformly distributed load, tip deflection is given by $\zeta = q \hat{L}^4 / 8 E J$, where $q$ is the load per unit length, $E$ is the Young's modulus, and $J = W h_f^3 / 12$ is the second moment of area . For our idealised beam, $\zeta = \sin (\gamma) \hat{L}$; and torque balance for load $q$ yields: $K \gamma = q \hat{L}^2 / 2$ leading to $q = 2 K \gamma / \hat{L}^2$. Substituting into the expression for $\zeta$ and assuming small deflection for which $\sin {\gamma} \simeq \gamma$, we get the following expression for the spring constant
\begin{equation}
K = \frac{4 E J}{\hat{L}} = \frac{E W h^3}{3 \hat{L}},
\end{equation}

To measure $E$, we performed mechanical tests as illustrated in  Figure~\ref{fw_fig:oscillograms}. Namely, we clamped the rigid part of each wing while letting the flexible part free, then subject the latter to an initial bending and observe its relaxation. Figure~\ref{fw_fig:oscillograms}(c-d) shows the response of the flexible parts of two wings characterized by $L_f/2 = 30$ mm, and $L_f/2 = 57.5$ mm to an initial bending $\zeta \simeq 15$ mm. From these oscillograms, one clearly see the natural frequency $f_n$ and the decay of the oscillatory motion. The natural frequency can be utilized to extract the Young's modulus $E$ using the following theoretical prediction from beam theory \cite{Reddy}:
\begin{equation}
f_n = \frac{1.875^2}{2 \pi} \sqrt{ \frac{E J}{\rho A \hat{L}^4} },
\end{equation}
where $A=h W$ designates the wing's cross section. For both wing types, we obtained $E = 3.6 \pm 0.5$ Gpa, in good agreement with values of the literature \cite{Szewczyk2006}. Additionally, we used the one-hinge model to predict the dynamics of the wing tip (blue line in Figure~\ref{fw_fig:oscillograms}c-d), and matched the decay of the oscillation amplitude for two lengths. Using the previously determined spring constant $K$ from the young's modulus $E$, and barring the adjustment of the damping coefficient $k_s = 4.5 \times 10^{-5} L_f/2$, the agreement between prediction and experiment is excellent. 


\begin{figure}[!htb]
\center
\includegraphics[width=0.99\textwidth]{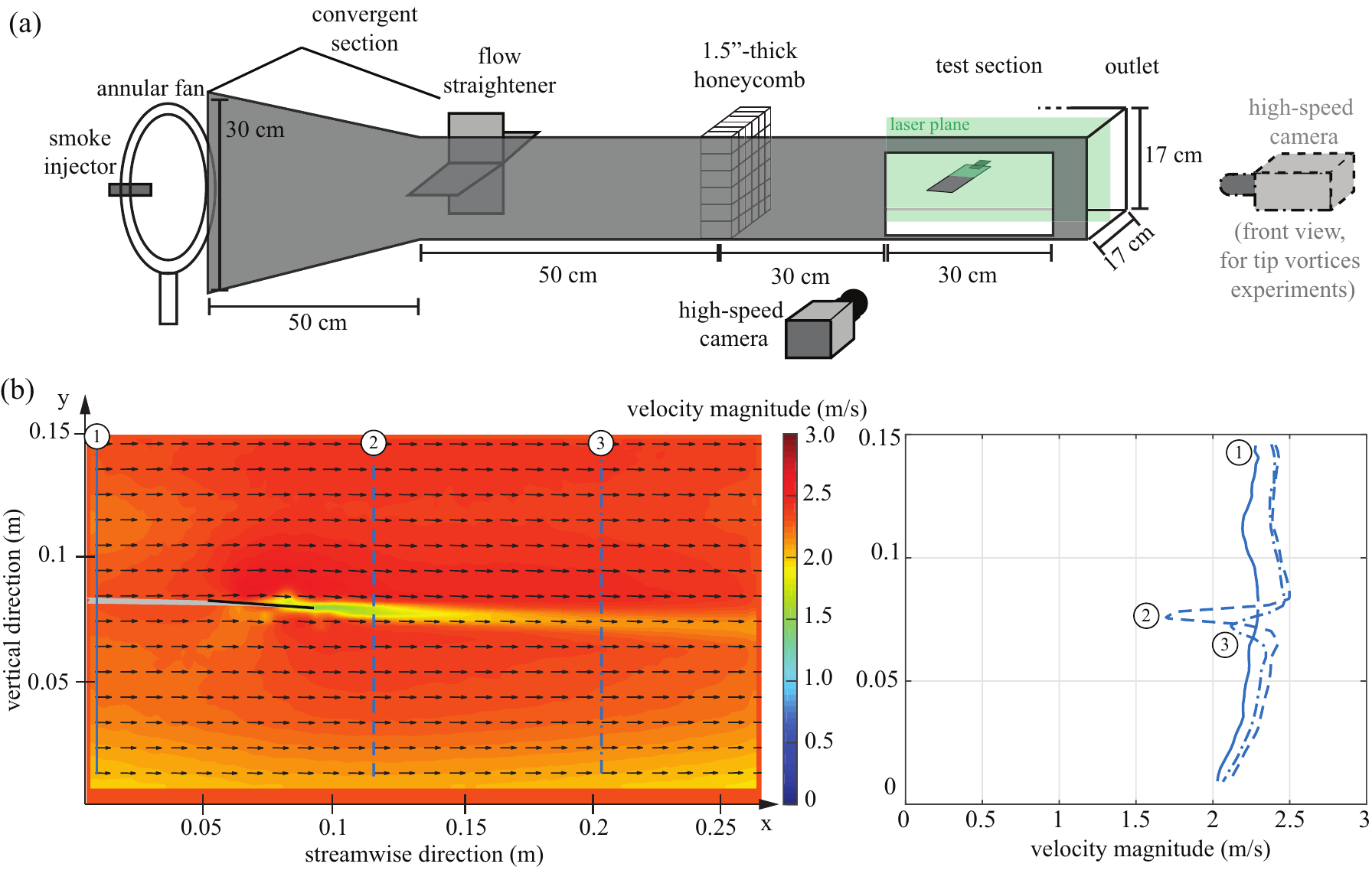}
\caption{\footnotesize{Low-speed wind tunnel. (a) global view of the setup. (b) Average streamwise velocity field ($\Delta t = 0.7$ s) at the wing's midspan obtained by Particle Image Velocimetry, and corresponding velocity profiles at various locations along the streamwise direction. The wing has a 5\degree angle of attack. The grey area shows the shadow cast by the wing, where the data is unavailable.}} 
\label{fig:wind_tunnel}
\end{figure}

 \section{Low-speed wind tunnel}

 We designed a low-speed wind tunnel with a $17 \times 17$ cm cross-section, see Figure~\ref{fig:wind_tunnel}(a). The air was supplied by an annular fan (Dyson, AM06) with controllable power from 1 to 10, allowing a mean velocity from $0.67$ m/s to $3.0$ m/s, respectively. For Particle Image Velocimetry (PIV) experiments, the air was seeded with a cloud of smoke generated by a commercial fog machine (1byone O00QL-0041) at the center of the fan. A high-speed camera equipped with a 60 mm macro lens was used to acquire PIV-ready images, which are processed using Matlab and the open-source app PIV lab \cite{ThielickeStamhuis2014}. The setup allowed both front and side optical access, through the outlet for front views, and a $30 \times 15$ cm plexiglass window for side views. The front view was used in the tip vortex experiments described in the main document, while the side view enable us to check the quality of the flow. Figure~\ref{fig:wind_tunnel}(b) shows a typical snapshot obtained with PIV (averaged over 120 individual snapshots, $\delta t = 0.7$ ms) of the flow over a half-wing (length $=80$ mm, chord $=42$ mm). Velocity profiles taken upstream the wing showed a reassuring, regular plug flow, and a gradually fading dip betraying the wing's presence for various downstream locations. Flow fields for other flow rates show similarly good characteristics.


\section{Scaling Analysis:  chordwise non-uniform wings}

We provide here predictions of the duration of the transient regime for chordwise-heterogeneous wings in the two possible regimes we identified in the main document: when linear momentum is dominant and when angular momentum is dominant.

When change in linear momentum is the limiting factor, we expect the duration of the transient regime to be set by the time needed to reach the terminal descent angle $\beta$, from an initial angle $\beta_\textrm{0}$. For our experiments, $\beta_\textrm{0} \approx 90 \degree$ and $\beta \approx 45 \degree$ , giving a target deflection $\Delta \beta = | \beta - \beta_\textrm{0} | \approx 45 \degree$, typical of tumbling wings~\cite{Vogel}. We now make the crude approximation that the deflection is solely due to the lift force $F_L \approx \frac{1}{2} \rho L W V_g^2$, perpendicular to the wing's trajectory by definition. The resulting circular trajectory has a radius $R = m V_g^2 / F_L = 2 (\rho_s / \rho) h $, and the target deflection is met after a trip of length $\ell = 2 \pi R ( \Delta \beta / 360 )$. Making the rough assumption that the velocity $V$ along this arc is constant and equal to $V_g$, we obtain the  (transient) travel duration: $\tau_\textrm{linear} = \ell / V_g \approx (\pi \Delta \beta / 180) \sqrt{ \rho_s / \rho } \sqrt{2 h/g} \approx 0.16$~s for our values for $\rho_s$, $\rho$, and $h$. This scaling law is independent of the non-dimensional moment of inertia $I / \rho L W^4$. 


Conversely, when the effect of moment of inertia is dominant, we expect the build-up of angular momentum to be the limiting factor. The duration of the transient regime is then set by the time needed to reach the terminal Strouhal number (non-dimensional tumbling rate) $\textrm{St} = \Omega W / 2 \pi / V_g$. For our experiments, $\textrm{St} \approx 0.15$, a value typical of tumbling wings \cite{AndersenWang2005,Mahadevan1999,Wang2013}). Ignoring the dissipation terms, the balance of angular momentum during the transient regime can be simply written as: $I \ddot{\theta} \approx  W F_L $, where the right-hand side corresponds to the torque generated by the lift force assuming a moment arm scaling as $W$. Replacing the time derivative by the associated time and angular velocity scales ($I \ddot{\theta} \approx \Omega / \tau_\textrm{angular}$), we obtain the transient duration:
$ \tau_\textrm{angular} = 4 \pi \textrm{St} (I/\rho L W^4) (W/V_g)$. 



A regime transition is expected when $\tau_\textrm{linear} \approx \tau_\textrm{angular}$, which corresponds to a non-dimensional moment of inertia $(I/\rho L W^4) \approx (\Delta \beta / 360) (\rho_s / \rho) (h/W) \textrm{St}^{-1} \simeq 4.0$, in good agreement with our experiments.



\bibliographystyle{vancouver}
\bibliography{Vincent2020}

\begin{thebibliography}{10}

\bibitem{LucasCostello2014}
Lucas KN, Johnson N, Beaulieu WT, Cathcart E, Tirrell G, Colin SP, et~al.
\newblock Bending rules for animal propulsion.
\newblock Nat Commun. 2014;5:3293.

\bibitem{Alben2008}
Alben S.
\newblock Optimal flexibility of a flapping appendage in an inviscid fluid.
\newblock J Fluid Mech. 2008;614:355--380.

\bibitem{Alben2012}
Alben S, Witt C, Baker TV, Anderson E, Lauder GV.
\newblock Dynamics of freely swimming flexible foils.
\newblock Phys Fluids. 2012;24:051901.

\bibitem{Quinn2015}
Quinn DB, Lauder GV, Smits AJ.
\newblock Maximizing the efficiency of a flexible propulsor using experimental
  optimization.
\newblock J Fluid Mech. 2015;767:430--448.

\bibitem{Heathcote2007}
Heathcote S, Gursul I.
\newblock Flexible flapping airfoil propulsion at low Reynolds numbers.
\newblock AIAA J. 2007;45(5):1066--1079.

\bibitem{Michelin2009}
Michelin S, Llewellyn-Smith SGL.
\newblock Resonance and propulsion performance of a heaving flexible wing.
\newblock Phys Fluids. 2009;21:071902.

\bibitem{RaspaRamananarivo2014}
Raspa S, Ramananarivo S, Thiria B, Godoy-Diana R.
\newblock Vortex-induced drag and the role of aspect ratio in undulatory
  swimmers.
\newblock Phys Fluids. 2014;26:041701.

\bibitem{Ramananarivo2010}
Ramananarivo S, Godoy-Diana R, Thiria B.
\newblock Rather than resonance, flapping wing flyers may play on aerodynamics
  to improve performance.
\newblock PNAS. 2010;108(15):5964--5969.

\bibitem{Lucas2015}
Lucas KN, Thornycroft PJM, Gemmell BJ, Colin SP, Costello JH, Lauder GV.
\newblock Effects of non-uniform stiffness on the swimming performance of a
  passively-flexing, fish-like foil model.
\newblock Bioinsp Biomim. 2015;10:056019.

\bibitem{Young2009}
Young J, Walker SM, Bomphrey RJ, Taylor GK, Thomas ALR.
\newblock Details of insect wing design and deformation enhance aerodynamic
  function and flight efficiency.
\newblock Science. 2009;325:1549--1552.

\bibitem{Le2010}
Le TQ, Ko JH, Byun D, Park SH, Park HC.
\newblock Effect of chord flexure on aerodynamic performance of a flapping
  wing.
\newblock J Bionic Eng. 2010;7:87--94.

\bibitem{Mittal2006}
Mittal R, Dong H, Bozkurttas M, Lauder GV, Madden P.
\newblock Locomotion with flexible propulsors: {II}. Computational modeling of
  pectoral fin swimming in sunfish.
\newblock Bioinsp Biomim. 2006;1:S35--S41.

\bibitem{LiuBose1997}
Liu P, Bose N.
\newblock Propulsive performance from oscillating propulsors with spanwise
  flexibility.
\newblock Proc R Soc Lond A. 1997;453:1763--1770.

\bibitem{Kang2011}
Kang CK, Aono H, Cesnik CES, Shyy W.
\newblock Effects of flexibility on the aerodynamic performance of flapping
  wings.
\newblock J Fluid Mech. 2011;689:32--74.

\bibitem{Heathcote2008}
Heathcote S, Wang Z, Gursul I.
\newblock Effect of spanwise flexibility on flapping wing propulsion.
\newblock J Fluid Struct. 2008;24:183--199.

\bibitem{Lentink2009}
Lentink D, Dickson WB, van Leeuwen JL, Dickinson MH.
\newblock Leading-Edge Vortices Elevate Lift of Leading-edge vortices elevate
  lift of autorotating plant seeds.
\newblock Science. 2009;324:1438.

\bibitem{Dupleich1941}
Dupleich P.
\newblock Rotation in free fall of rectangular wings of elongated shape.
\newblock NACA Technical Memo; 1941. 1201.

\bibitem{Belmonte1998}
Belmonte A, Eisenberg H, Moses E.
\newblock From Flutter to Tumble: Inertial Drag and Froude Similarity in
  Falling Paper.
\newblock Phy Rev Lett. 1998;81(2):345--348.

\bibitem{Mahadevan1999}
Mahadevan L, Ryu WS, Samuel ADT.
\newblock Tumbling cards.
\newblock Phys Fluids. 1999;11(1):1--3.

\bibitem{PesaventoWang2004}
Pesavento U, Wang ZJ.
\newblock Falling paper: Navier-Stokes solutions, model of fluid forces, and
  center of mass elevation.
\newblock Phys Rev Lett. 2004;93:144501.

\bibitem{AndersenWang2005}
Andersen A, Pesavento U, Wang ZJ.
\newblock Unsteady aerodynamics of fluttering and tumbling plates.
\newblock J Fluid Mech. 2005;541:65--90.

\bibitem{TamBushKudrolli2010}
Tam D, Bush JWM, Robitaille M, Kudrolli A.
\newblock Tumbling dynamics of passive flexible wings.
\newblock Phys Rev Lett. 2010;104:184504.

\bibitem{Tam2015}
Tam D.
\newblock Flexibility increases lift for passive fluttering wings.
\newblock J Fluid Mech. 2015;765:R2.

\bibitem{Anderson2001}
Anderson JD.
\newblock Fundamentals of Aerodynamics.
\newblock Third edition ed. McGraw-Hill Higher Education; 2001.

\bibitem{AlbenShelleyZhang2002}
Alben S, Shelley M, Zhang J.
\newblock Drag reduction through self-similar bending of a flexible body.
\newblock Nature. 2002;420:479--481.

\bibitem{AlbenShelleyZhang2004}
Alben S, Shelley M, Zhang J.
\newblock How flexibility induces streamlining in a two-dimensional flow.
\newblock Phys Fluids. 2004;16(5):1694--1713.

\bibitem{DeLangre2008}
de~Langre E.
\newblock Effects of Wind on Plants.
\newblock Annu Rev Fluid Mech. 2008 seed dispersal;40:141--168.

\bibitem{Gosselin2010}
Gosselin F, de~Langre E, Machado-Almeida BA.
\newblock Drag reduction of flexible plates by reconfiguration.
\newblock J Fluid Mech. 2010;650:319--341.

\bibitem{Hoerner1965}
Hoerner SF.
\newblock Fluid-Dynamic Drag.
\newblock S. F. Hoerner; 1965.

\bibitem{Dommasch1967}
Dommasch DO, Sherby SS, Connolly TF.
\newblock Airplane Aerodynamics.
\newblock Fourth edition ed. Pitman Publishing Corporation; 1967.

\bibitem{Green1995}
Green SI, editor.
\newblock Fluid Vortices: Fluid Mechanics and Its Applications.
\newblock Kluwer Academic Publishers; 1995.

\bibitem{BirchDickinson2001}
Birch JM, Dickinson MH.
\newblock Spanwise flow and the attachment of the leading-edge vortex on insect
  wings.
\newblock Nature. 2001;412:729--732.

\bibitem{Spedding1992}
Spedding GR.
\newblock 3: The Aerodynamics of Flight.
\newblock In: Advances in Comparative and Environmental Physiology.
  Springer-Verlag, Berlin; 1992. .

\bibitem{Schmitz1941}
Schmitz FW.
\newblock Aerodynamics of the model airplane. Part I. Airfoil Measurements.
\newblock NASA; 1941. RSIC-721.

\bibitem{Kowarik2007}
Kowarik I, S\"aumel I.
\newblock Biological flora of Central Europe: \textit{Ailanthus altissima}
  (Mill.) Swingle.
\newblock Perspect plant ecol. 2007;7:207--237.

\bibitem{Kim2013}
Kim S, Laschi C, Trimmer B.
\newblock Soft robotics: a bioinspired evolution in robotics.
\newblock Trends in Biotechnology. 2013;31(5):287 -- 294.
\newblock Available from:
  \url{http://www.sciencedirect.com/science/article/pii/S0167779913000632}.

\bibitem{Tolley2014}
Tolley MT, Shepherd RF, Mosadegh B, Galloway KC, Wehner M, Karpelson M, et~al.
\newblock {A} resilient, {U}ntethered {S}oft {R}obot.
\newblock Soft Robotics. 2014;1(3):213--223.

\bibitem{Rich2018}
Rich SI, Wood RJ, Majidi C.
\newblock Untethered soft robotics.
\newblock Nature Electronics. 2018;1:102--112.

\bibitem{Nawroth2012}
Nawroth JC, Lee H, Feinberg AW, Ripplinger CM, McCain ML, Grosberg A, et~al.
\newblock A tissue-engineered jellyfish with biomimetic propulsion.
\newblock Nat Biotechnol. 2012;30(8):792--797.

\bibitem{ParkParker2016}
Park SJ, Gazzola M, Park KS, Park S, Santo VD, Blevins EL, et~al.
\newblock Phototactic guidance of a tissue-engineered soft-robotic ray.
\newblock Science. 2016;353:158--162.

\bibitem{Colin2012}
Colin SP, Costello JH, Dabiri JO, Villanueva A, Blottman JB, Gemmell BJ, et~al.
\newblock Biomimetic and Live Medusae Reveal the Mechanistic Advantages of a
  Flexible Bell Margin.
\newblock PLOS ONE. 2012 11;7(11):1--10.
\newblock Available from: \url{https://doi.org/10.1371/journal.pone.0048909}.

\bibitem{PelletierMueller2000}
Pelletier A, Mueller TJ.
\newblock Low Reynolds Number Aerodynamics of Low-Aspect-Ratio,
  Thin/Flat/Cambered-Plate Wings.
\newblock J Aircraft. 2000;37(5):825--832.

\bibitem{Ortiz2015}
Ortiz X, Rival D, Wood D.
\newblock Forces and Moments on Flat Plates of Small Aspect Ratio with
  Application to PV Wind Loads and Small Wind Turbine Blades.
\newblock Energies. 2015;8:2438--2453.

\bibitem{Laitone1997}
Laitone EV.
\newblock Wind tunnel tests of wings at Reynolds numbers below 70 000.
\newblock Exp Fluids. 1997;23:405--409.

\bibitem{Reddy}
Reddy JN.
\newblock Theory and analysis of elastic plates and shells.
\newblock 2nd ed. CRC Press; 2006.

\bibitem{Szewczyk2006}
Szewczyk W, Marynowski K, Tarnawski W.
\newblock An analysis of young's modulus distribution in the paper plane.
\newblock Fibres and Textiles in Eastern Europe. 2006;14(4):91--94.

\bibitem{ThielickeStamhuis2014}
Thielicke W, Stamhuis EJ.
\newblock PIVLab - Towards User-friendly, Affordable and Accurate Digital
  Particle Image Velocimetry in MATLAB.
\newblock JORS. 2014;2(1):e30.

\bibitem{Vogel}
Vogel S.
\newblock Life in moving fluids: the physical biology of flow.
\newblock Princeton University Press; 1994.

\bibitem{Wang2013}
Wang WB, Hu RF, Xu SJ, Wu ZN.
\newblock Influence of aspect ratio on tumbling plates.
\newblock J Fluid Mech. 2013;733:650--679.

\end{thebibliography}

\end{document}